\documentclass[twocolumn,preprintnumbers,prl,superscriptaddress,longbibliography]{revtex4-1}

\usepackage{graphicx}
\usepackage{dcolumn}
\usepackage{bm}

\usepackage[colorlinks=true,linkcolor=black, citecolor=black,
urlcolor=black]{hyperref}

\usepackage{relsize}
 \usepackage{multirow,graphics}
 \usepackage{amstext}
 \usepackage{amssymb}
 \usepackage{amsmath}
 \usepackage{graphicx}
 \usepackage{color}
 \usepackage{dsfont} 
 \usepackage{delimset} 
 \usepackage{multirow} 

\newcommand{\sfrac}[2]{{\textstyle\frac{#1}{#2}}}
\newcommand{\half}{\sfrac{1}{2}}
\newcommand{\ihalf}{\sfrac{i}{2}}

\newcommand{\gen}[1]{\mathrm{#1}}
\newcommand{\Eval}{s} 
\newcommand{\fd}{{\text{extra}}}

\newcommand{\levo}[1]{\mathrm{\widehat #1}}

\newcommand{\dd}{\mathrm{d}}

\makeatletter
\newlength{\apb@width}
\newcommand{\autoparbox}[2][c]{\settowidth{\apb@width}{#2}\parbox[#1]{\apb@width}{#2}}
\newcommand{\includegraphicsbox}[2][]{\autoparbox{\includegraphics[#1]{#2}}}
\makeatother

\makeatletter
\def\mr@ignsp#1 {\ifx\:#1\@empty\else #1\expandafter\mr@ignsp\fi}%
\newcommand{\multiref}[1]{\begingroup
\xdef\mr@no@sparg{\expandafter\mr@ignsp#1 \: }%
\def\mr@comma{}%
\@for\mr@refs:=\mr@no@sparg\do{\mr@comma\def\mr@comma{,}\ref{\mr@refs}}%
\endgroup}
\makeatother

\newcommand{\hypref}[2]{\ifx\href\asklfhas #2\else\href{#1}{#2}\fi}

\newcommand{\Tabref}[1]{table~\multiref{#1}}

\renewcommand{\eqref}[1]{(\multiref{#1})}

\begin{document}

\preprint{HU-EP-20/11}

\title{Massive Conformal Symmetry and Integrability for Feynman Integrals}

%
\author{Florian Loebbert}
\email{loebbert@physik.hu-berlin.de} 
\affiliation{%
Institut f\"ur Physik, Humboldt-Universi\"at zu Berlin,
Zum Gro{\ss}en Windkanal 6, 12489 Berlin, Germany
}%
\author{Julian Miczajka} 
\email{miczajka@physik.hu-berlin.de}
\affiliation{%
Institut f\"ur Physik, Humboldt-Universi\"at zu Berlin,
Zum Gro{\ss}en Windkanal 6, 12489 Berlin, Germany
}%
\author{Dennis M\"uller} 
\email{dennis.mueller@nbi.ku.dk}
\affiliation{%
Niels Bohr Institute, Copenhagen University, Blegdamsvej 17, 
2100 Copenhagen, Denmark}%
\author{Hagen M\"unkler} 
\email{muenkler@itp.phys.ethz.ch}
\affiliation{%
Institut f\"ur Theoretische Physik,
Eidgen\"ossische Technische Hochschule Z\"urich,
Wolfgang-Pauli-Strasse 27, 8093 Z\"urich, Switzerland}%



\date{\today}

\begin{abstract}

In the context of planar holography, integrability plays an important role for solving certain massless quantum field theories such as $\mathcal{N}=4$ SYM theory. In this letter we show that integrability also features in the building blocks of massive quantum field theories.
 At one-loop order we prove that all massive $n$-gon  Feynman integrals in generic spacetime dimensions are invariant under a massive Yangian symmetry. At two loops similar statements can be proven for graphs built from two $n$-gons. At generic loop order we conjecture that all graphs cut from regular tilings of the plane with massive propagators on the boundary are invariant. We support this conjecture by a number of numerical tests for higher loops and legs.
 The observed Yangian extends the bosonic part of the massive dual conformal symmetry that was found a decade ago on the Coulomb branch of $\mathcal{N}=4$ SYM theory. By translating the Yangian level-one generators from dual to original momentum space, we introduce a massive generalization of momentum space conformal symmetry. Even for non-dual conformal integrals this novel symmetry persists. The Yangian can thus be understood as the closure of massive dual conformal symmetry and this new massive momentum space conformal symmetry, which suggests an interpretation via AdS/CFT. As an application of our findings, we bootstrap the hypergeometric building blocks for examples of massive Feynman integrals. 
%

%
\end{abstract}

                              
\maketitle

\section{Introduction}
\label{sec:intro}

In 1954 Wick and Cutkosky noticed that a certain class of ladder-type Feynman integrals with massive propagators features a massive dual conformal symmetry \cite{Wick:1954eu,Cutkosky:1954ru}. 
While most of the formal insights into quantum field theory inspired by the AdS/CFT correspondence are limited to massless situations, this massive dual conformal symmetry is naturally realized in the context of this duality~\cite{Alday:2009zm,Caron-Huot:2014gia}. In particular, the extended dual conformal symmetry limits the variables that certain massive Feynman integrals can depend on and thus simplifies their computation. In the present letter we argue that for large classes of Feynman integrals, this massive dual conformal symmetry is in fact only the zeroth level of an infinite dimensional massive Yangian algebra. In addition to limiting the number of variables, this new symmetry strongly constrains the functional form of the integrals. While these symmetry properties naturally extend the observations on the integrability of massless Feynman integrals \cite{Chicherin:2017cns,Chicherin:2017frs,Loebbert:2019vcj}, to the knowledge of the authors this is the first occurence of quantum integrability in massive quantum field theory in $D>2$ spacetime dimensions. 

For \emph{massless} $\mathcal{N}=4$ super Yang--Mills (SYM) theory it was recently argued that planar integrability is preserved in a certain double scaling limit, which (in the simplest case) results in the so-called bi-scalar fishnet theory \cite{Gurdogan:2015csr}. Here, individual (massless) Feynman integrals inherit the Yangian symmetry that underlies the integrability of the prototypical examples of the AdS/CFT duality \cite{Chicherin:2017cns,Chicherin:2017frs}. A similar starting point, i.e.\ an \emph{integrable massive} avatar of $\mathcal{N}=4$ super Yang--Mills theory, is not known. We thus investigate massive Feynman integrals directly, i.e.\ we consider the properties of functions of the type
\begin{equation}
\includegraphicsbox{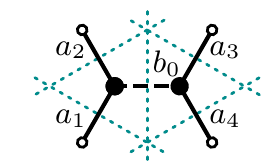}
\quad
=
\int \frac{\dd^D x_0 \dd^D x_{\bar 0}}
{
\hat x_{01}^{2a_1}
\hat x_{02}^{2a_2}
x_{0\bar 0}^{2b_0}
\hat x_{\bar 03}^{2a_3}
\hat x_{\bar 04}^{2a_4}
},
\end{equation}
where $x_{jk}^\mu=x_j^\mu-x_k^\mu$ and $\hat x_{jk}^2=x_{jk}^2+(m_j-m_k)^2$. Here the dashed internal propagator is massless, i.e.\ $m_0=m_{\bar 0}=0$, while the other propagators are massive.
The $x$-variables denote dualized momenta (dotted green diagram) related via $p^{\mu}_j=x^{\mu}_j-x^{\mu}_{j+1}$  \footnote{Note that the $p_j^2$ are unconstrained and the $m_j$ are generic; we have $x^{\mu}_j=x^{\mu}_{1}-\sum_{k < j} p^{\mu}_k $ and $x_{n+1}^\mu=x_{1}^\mu$.}.
Our findings suggest that all Feynman graphs, which are cut from regular tilings of the plane and have massive propagators on the boundary, feature a massive $D$-dimensional Yangian symmetry. 
We will demonstrate the usefulness of this Yangian for bootstrapping massive Feynman integrals. Finally, we will show that, when translated to momentum space, the non-local Yangian symmetry can be interpreted as a massive generalization of momentum space conformal symmetry. 
This suggests to interpret this novel symmetry within the AdS/CFT correspondence.

%
\section{Massive Yangian}

Massive dual conformal symmetry is realized in the form of partial differential equations obeyed by coordinate space Feynman integrals. That is, the integrals are annihilated by the tensor product action of the level-zero dual conformal generators
$
\gen{J}^a = \sum_{j=1}^n \gen{J}_{j}^a, 
$
where $\gen{J}_j^a$
denotes one of 
the following densities  acting on $x_j$:
\begin{align}
\gen{P}^{\hat \mu}_j &= -i \partial_{x_{j}}^{\hat \mu}, 
\qquad\qquad
\gen{L}_j^{\hat \mu\hat \nu} = i x_j^{\hat \mu} \partial_{x_{j}}^{\hat \nu} - ix^{\hat \nu}_j \partial_{x_{j}}^{\hat \mu}, 
\nonumber
\\
\gen{D}_j &= -i \brk!{x_{j\mu} \partial_{x_j}^\mu + m_j \partial_{m_j} + \Delta_j},
\label{eqn:massdualconfrep}
\\
 \gen{K}^{\hat \mu}_j &= -2ix_j^{\hat \mu}\brk!{x_{j\nu}  \partial_{x_j}^\nu  + m_j\partial_{m_j} + \Delta_j} +i (x^2_j + m^2_j)\partial_{x_{j}}^{\hat \mu}.\notag
\end{align}
These can be understood as massless generators in $D+1$ dimensions with $x_j^{D+1}=m_j$. Only the components $\hat \mu=1,\dots,D$ of the generators correspond to symmetries.
Here we work with the Euclidean metric and the index $\hat{\mu}$ runs from 1~to~$D+1$, while $\mu$ runs from 1~to~$D$.

The massive Yangian is spanned by the above level-zero Lie algebra generators and the bi-local level-one generators 
defined as
\begin{equation}
\label{eq:DefLev1}
 \gen{\widehat J}^a=\half f^a{}_{bc}\sum_{j<k=1}^n \gen{J}_{j}^c \gen{J}_{k}^b+ \sum_{j=1}^n \Eval_j \gen{J}_{j}^a,
\end{equation}
where $ f^a{}_{bc}$ denotes the Lie algebra structure constants.
The so-called evaluation parameters $\Eval_j$ depend on the considered Feynman integral and will be specified below.
The level-one momentum generator for instance reads 
\begin{align}
\label{eq:DefPhat}
\gen{\widehat P}^{\hat\mu}
=\sfrac{i}{2} \sum_{j,k=1}^n 
\text{sign}(k-j)\brk!{\gen{P}_j^{\hat\mu} \gen{D}_k + \gen{P}_{j\nu} \gen{L}_k^{\hat\mu \nu} }
 + \sum_{j=1}^n \Eval_j \gen{P}_j^{\hat \mu}.
\end{align}
Here we do not sum the internal index $\nu$ over $D+1$ dimensions but rather define this contribution separately, 
e.g.\
$\gen{\widehat P}^{\hat \mu}_\fd=\sfrac{i}{2}\sum_{j,k}\text{sign}(k-j)\gen{P}_{j,D+1} \gen{L}_k^{\hat\mu,D+1}$.
Note that the whole Yangian algebra is spanned by the above level-one momentum generator \eqref{eq:DefPhat} and the level-zero generators~\eqref{eqn:massdualconfrep}. 

\section{One and Two Loops }

First, we consider a generic scalar $n$-point Feynman integral at one-loop order with massive propagators. Propagator powers $a_j$ and the spacetime dimension $D$ are arbitrary:
\begin{align}
I_n = \int \text{d}^D x_0 \ \rho_n
=\includegraphicsbox{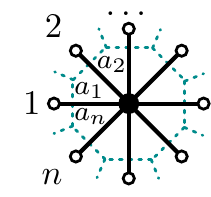}\,.
\label{eq:nGonIntegral}
\end{align}
Here $\rho_n
=
\prod_{j=1}^n (x_{0j}^2 + m_j^2)^{-a_j}$ and we use the notation $x_{jk}^\mu=x_j^\mu-x_k^\mu$.
These integrals have been expressed in terms of hypergeometric functions by Davydychev~\cite{Davydychev:1990cq}.
Surprisingly, we find that the integrand is invariant under the massive level-one generators acting on legs $1$ to $n$ defined in \eqref{eq:DefLev1}, 
i.e.\ we have $\gen{\widehat J}^a \rho_n = 0$ which implies 
\begin{align}
\gen{\widehat J}^a I_n = 0.
\label{eq:Lvl1Symm}
\end{align}
The scaling dimensions in \eqref{eqn:massdualconfrep} take values $\Delta_j=a_j$ and the evaluation parameters entering the level-one generators \eqref{eq:DefLev1} can be written in the compact form
\begin{align}
\Eval_j
=\half \sum_{k=1}^n \text{sign}(k-j) \,a_k.
\label{eq:EvalsNGons}
\end{align}
We emphasize that this symmetry also holds for cases where an arbitrary subset of the masses is set to zero~
\footnote{In these cases one acts on massive legs with the generators of the massive symmetry and on massless legs with the usual dual-conformal generators obtained by dropping the mass terms in \protect\eqref{eqn:massdualconfrep}.}.
Moreover, the above level-one symmetry holds even if we do not impose conformal level-zero symmetry, i.e.\ the constraint $\sum_{j=1}^n a_j=D$ on the propagator powers. Hence, the level-one symmetry can be used to constrain dual conformal and non-dual-conformal integrals.
Finally we should mention that also the generators as $\gen{\widehat P}^{\hat \mu}_\fd$ given below \eqref{eq:DefPhat}, which can be interpreted as contributions to an internal sum over $D+1$ dimensions, define a separate symmetry of the above one-loop integrands \footnote{Switching on the internal mass $m_0$, only the combination $\gen{\widehat P}^{\hat \mu}+\gen{\widehat P}^{\hat \mu}_\fd$ is a symmetry.}.

\begin{table}
\renewcommand{\arraystretch}{1.3}
\begin{tabular}{|l|c|c|c|}\hline
Loops&Graphs&Conformal&Not Conformal
\\\hline
One&$n$-gons&Full Yangian \& $\levo{J}_\fd^a$ & All $\levo{J}^a$  \& $\levo{J}_\fd^a$ 
\\
Two&$l$-$r$-gons&Full Yangian&  $\levo{P}^\mu$
\\
All&Tilings&Full Yangian&  $\levo{P}^\mu$
\\\hline
\end{tabular}
\caption{Overview of non-local symmetries. We speak of a (dual) conformal integral, if at each $x$-space vertex the propagator powers $a_j$ sum up to the spacetime dimension $D$.}
\label{tab:allsyms}
\end{table}

While \eqref{eq:Lvl1Symm} represents the central symmetry equation that extends to higher loop integrals, see \Tabref{tab:allsyms}, at one-loop order even stronger invariance statements can be formulated:
Firstly, the one-loop invariance even holds at the integrand level. Secondly, due to the total permutation symmetry, the $n$-gon integrand is even invariant under the two-point generator density:
\begin{equation}
\label{eq:TwoPointSymNgon}
\gen{\widehat J}^a_{jk}\rho_n=0,
\qquad
\gen{\widehat J}^a_{jk}=\half f^a{}_{bc} \gen{J}_{j}^c \gen{J}_{k}^b+ \Eval_j \gen{J}_{j}^a+ \Eval_k \gen{J}_{k}^a,
\end{equation}
for $j,k\in\{1,\dots,n\}$ and with  
$\Eval_j= a_k/2$ and
$\Eval_k=- a_j/2$.

At two loops we find that $n$-point integrals (not integrands) built from an $l+1$-gon and an $r+1$-gon with $n=l+r$ connected by a massless internal propagator are Yangian invariant:
\begin{equation}
\includegraphicsbox{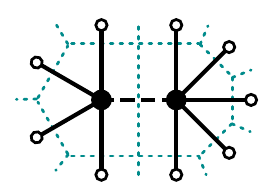}
\end{equation}
Using the permutation symmetries of this class of diagrams as well as their level-zero symmetry and the two-point level-one symmetry \eqref{eq:TwoPointSymNgon} of the one-loop integrals, it is not difficult to prove this two-loop invariance.

\section{Higher Loops}
\label{sec:HigherLoops}

Beyond the analytic proofs for one- and two-loop integrals, we have numerically tested the invariance of higher loop diagrams for various examples, cf.\ \Tabref{tab:ExampleGraphs}.
\begin{table}
\renewcommand{\arraystretch}{1.1}
\begin{tabular}{|c | c | c||c|}\hline
\multicolumn{3}{|c||}{Non-local Symmetry}&{No Sym.}
\\\hline
Triangle&Square&Hexagon&Irregular\\
\hline
\includegraphicsbox[scale=.7]{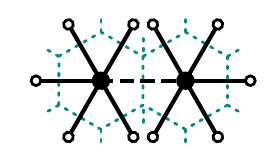}
&
\includegraphicsbox[scale=.7]{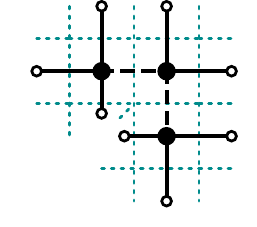}
&
\includegraphicsbox[scale=.7]{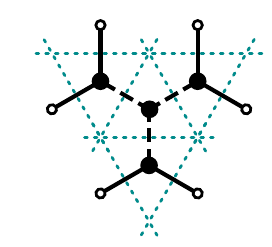}
&
\includegraphicsbox[scale=.6]{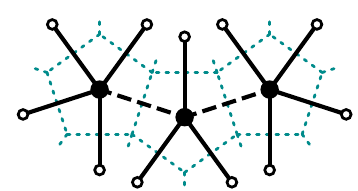}
\\
\includegraphicsbox[scale=.7]{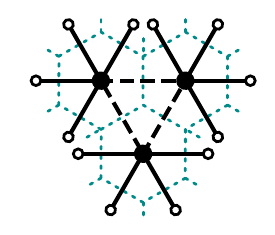}
&
\includegraphicsbox[scale=.7]{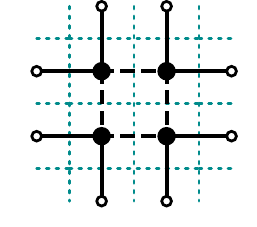}
&
\raisebox{1mm}{\includegraphicsbox[scale=.7]{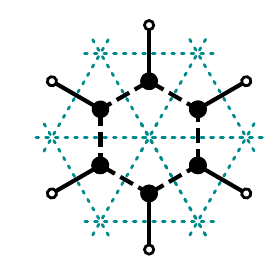}}
&
\includegraphicsbox[scale=.7]{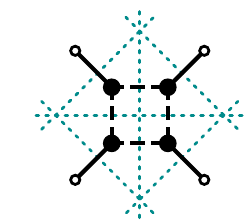}
\\\hline
\end{tabular}
\caption{
Examples of integrals up to 6 loops and 12 legs explicitly tested for level-one momentum invariance by acting on the Feynman parametrization. Graphs cut from regular tilings are invariant if propagators on the boundary of the graph are massive or massless (solid), while internal propagators have to be massless (dashed). Graphs not cut from regular tilings have no level-one symmetry as described here.}
\label{tab:ExampleGraphs}
\end{table}
To be precise, we have checked for the presence of the symmetry by acting on the respective Feynman parametrization and numerical integration of the result.
On this basis we conjecture that all planar Feynman graphs, which are cut along a closed contour from one of the three regular tilings of the plane, have massive Yangian symmetry if all internal propagators are massless. The external 
propagators can be massive or massless.
Here an $x$-space propagator is called external if it is connected to an external point of the diagram. The integration can be evaluated in generic spacetime dimension $D$ and propagator powers may take generic values $a_j$.
For full Yangian invariance we require the conformal constraint
$D=\sum_{j\in \text{vertex}} a_j$ at each integration vertex.
If this constraint is not satisfied, we still have level-one momentum symmetry $\levo{P}^\mu$, which corresponds to a massive generalization of momentum space conformal symmetry that we introduce below.
In the massless limit this matches the observations of \cite{Chicherin:2017cns,Chicherin:2017frs,Loebbert:2019vcj} on the integrability of massless Feynman integrals in $D=3$, 4 and 6 spacetime dimensions.

The values of the evaluation parameters $\Eval_j$ entering \eqref{eq:DefPhat} can be read off from the Feynman graph as follows: 
i)~Choose an arbitrary leg $1$ with $s_1=\sfrac{1}{2}\sum_{k=2}^n a_k$, ii)~walk clockwise along the external boundary of the ($x$-space) graph and pick up the respective propagator weights, iii)~here an external propagator with power $a_j$ contributes $-{a_j}/{2}$ while an internal propagator with weight $b_k$ contributes $-b_k + {D}/{2}$. The evaluation parameter $\Eval_{j+1}$ is obtained from $\Eval_j$ by adding the respective terms, e.g.\
\begin{equation}
\label{eq:EvalRules}
\includegraphicsbox{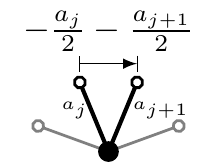}
\qquad
\includegraphicsbox{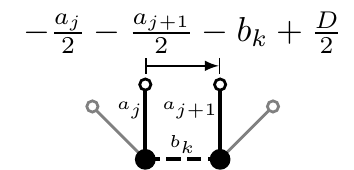}.
\end{equation}
\section{Yangian Bootstrap}


Let us now employ the above non-local symmetry to bootstrap Feynman integrals with massive propagators. This discussion extends the algorithm of \cite{Loebbert:2019vcj} to the massive case.
We start with some examples containing conformal vertices obeying the condition $\sum_j a_j=D$. Notably, after solving these integrals, we can obtain an infinite class of integrals by acting with $r$ mass derivatives $\partial_{m_k}$ which yields an integral with propagator weights $\tilde a_j$ with $\sum_j \tilde a_j=D+r$.

\paragraph{Conformal, 2 points, 2 masses, 1 loop:}

As a simple starting example consider the two-point integral 
(see e.g.\ \cite{Boos:1990rg,Davydychev:1990cq,Kniehl:2011ym} for the non-conformal case)
\begin{align}
I_2^{m_1m_2} = 
\includegraphicsbox{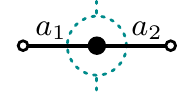}
=\frac{ (1-v^2)^{\beta/2}}{m_1^{a_1} m_2^{a_2}}\phi(v),
\end{align}
written in terms of a function $\phi$ of the conformally invariant variable
$v 
=(m_1^2+m_2^2+x_{12}^2)/2m_1m_2$.
For convenience we set
$2\alpha=a_1-a_2-1$
and 
$2\beta=-a_1-a_2+1$.
Acting on $I_2$ with the level-one momentum generator yields the associated Legendre differential equation \footnote{Note that this equation arises from the differential operator $\levo{ P}^{\hat \mu}_\fd$, while the rest of $\levo{P}$ trivially annihilates $I_2$. This is a generic feature for two-point invariants.}:
\begin{equation}
\brk[s]*{\alpha(\alpha+1)+\sfrac{\beta^2}{v^2-1}}\phi-2v\phi'+(1-v^2)\phi''=0.
\end{equation}
The general solution to the above differential equation is easily found, e.g.\ in Mathematica, to be a linear combination of the associated Legendre functions of the first and second kind, i.e.  $P_\alpha^\beta$ and $Q_\alpha^\beta$. The coefficients of these functions can be fixed using numerical input for the integral, which yields
\begin{equation}
\phi(v) = 2^{\beta}\pi^{1-\beta} P_\alpha^\beta(v).
\end{equation}
We have thus completely constrained the integral.
\paragraph{Conformal, 3 points, 3 masses, 1 loop:}
Consider next the one-loop three-point integral with all propagators massive and the conformal constraint $D=a_1+a_2+a_3$ (see e.g.\ \cite{Nickel:1978ds,Davydychev:2017bbl,Bourjaily:2019exo} for the case with $a_j=1$):
\begin{equation}
\label{eq:I3start}
I_3^{m_1m_2m_3}=
\includegraphicsbox{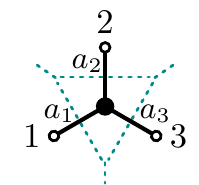}=\frac{\phi_3(u,v,w)}{m_1^{a_1}m_2^{a_2}m_3^{a_3}}.
\end{equation}
Here we have chosen the conformal variables $u=u_{12}$, $v=u_{13}$ and $w=u_{23}$, where
$u_{jk}=
-\hat x_{jk}^2/4m_jm_k$.
Making a series ansatz in $u$, $v$ and $w$, the Yangian PDEs translate into recurrence equations for the series coefficients. 
These equations can straightforwardly be solved, which yields Srivastava's triple hypergeometric function $H_{C}$, a generalization of Appell's hypergeometric function $F_1$, see e.g.\ \cite{srivastava2015note}:
\begin{equation}
\label{eq:SrivastavaHC}
H_{C}(u,v,w)
=
\sum_{k,l,n=0}^\infty \frac{(a_1)_{k+l}(a_2)_{k+n}(a_3)_{l+n}}{(\gamma)_{k+l+n}}\frac{u^k}{k!}\frac{v^l}{l!}\frac{w^n}{n!},
\end{equation}
with $\gamma=D/2+1/2$ and 
the Pochhammer symbol
$(a)_k= \Gamma_{a+k}/\Gamma_a.$
This series converges for
\begin{equation}
\label{eq:ConvergenceHC}
 \abs{u}+\abs{v}+\abs{w}-2\sqrt{(1-\abs{u})(1-\abs{v})(1-\abs{w})}<2,
\end{equation}
and in this region it is the only converging solution of 29 formal series solutions to the Yangian equations. Numerical comparison to the Feynman parameter representation of 
\eqref{eq:I3start} yields the overall constant such that
\begin{equation}
I_3^{m_1m_2m_3}
=
\frac{\pi^{D/2}\Gamma_{D/2}}{\Gamma_D m_1^{a_1}m_2^{a_2}m_3^{a_3}}
H_{C}(u,v,w).
\end{equation}
To the knowledge of the authors this result has not been given elsewhere in the literature.


\paragraph{Non-Conformal, 3 points, 1 mass, 1 loop:}

Let us finally demonstrate the usefulness of the level-one symmetry 
for an integral that has no massive dual conformal (level-zero) symmetry. We write the triangle with massive leg $1$ as
\begin{equation}
I_3^{m_100} = 
\includegraphicsbox{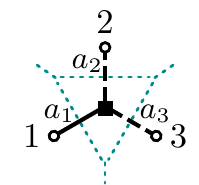}
=m_1^{D-2a_1-2a_2-2a_3} \psi_3\left(u,v,w\right),
\end{equation}
with
$u=u_{12}$,
$v=u_{13}$ and
$w=-u_{23}$,
and $u_{jk}=-x_{jk}^2/m_1^2$.
We act with the level-one PDEs on the series ansatz
\begin{equation}
\label{eq:SeriesAnsatz3pt1Mass}
G_{xyz} ^{\alpha_1 \alpha_2 \alpha_3\gamma_1\gamma_2} 
	= \sum_{k,l,n}
		f_{kln} \,\frac{u^k}{k!} \frac{v^l}{l!} \frac{w^n}{n!},
\end{equation}
where the sum runs over $k\in x+\mathbb{Z}$, $l\in y+\mathbb{Z}$, $n\in z+\mathbb{Z}$ for some $(x,y,z)\in \mathbb{R}^3$.
The resulting recurrence equations can straightforwardly be solved which results in
\begin{align}
f_{kln}
=
 \frac{(\alpha_1)_{k+l+n}(\alpha_2)_{k+n}(\alpha_3)_{l+n}}{(\gamma_1)_{k+l+n}(\gamma_2)_{n}},
\label{eq:FundSol3GonNotConf1Mass}
\end{align}
with the new parameters
\begin{equation}
\label{eq:params3pt1MassNonConf} 
\alpha_1 = a_1 + a_2 + a_3-\sfrac{D}{2}   , \quad
\alpha_2 =  a_2 , \quad
\alpha_3 =   a_3,  \quad
\gamma_1 = \sfrac{D}{2},
\end{equation}
and the abbreviation
$\gamma_2 = 1 - \gamma_1 + \alpha_2 +\alpha_3$.
We find 36 combinations of $(x,y,z)$ in \eqref{eq:SeriesAnsatz3pt1Mass} for which the series terminates after rescaling, which
we take as a necessary criterion for convergence. 
However, only for the two choices $(0,0,0)$ and 
$(0,0,1-\gamma_2)$ of these 36 combinations, $u$, $v$ and $w$ are the effective series variables.
In fact, the ansatz combining these two functions, 
i.e.\ 
\footnote{
Note the following shift identity for $A=G_{001-\gamma_2} ^{\alpha_1 \alpha_2 \alpha_3\gamma_1\gamma_2} 
$
and 
$B=G_{000}^{\alpha_1-\gamma_2+1,\alpha_2-\gamma_2+1,\alpha_3-\gamma_2+1,\gamma_1-\gamma_2+1,2-\gamma_2}$:
\begin{equation*}
A
=
w^{1-\gamma_2} \frac{\Gamma_{\gamma_1}\Gamma_{\gamma_2}\Gamma_{1+\alpha_1-\gamma_2}\Gamma_{1+\alpha_2-\gamma_2}\Gamma_{1+\alpha_3-\gamma_2}}{\Gamma_{\alpha_1}\Gamma_{\alpha_2}\Gamma_{\alpha_3}\Gamma_{2-\gamma_2}\Gamma_{1+\gamma_1-\gamma_2}} B
\end{equation*}
}
\begin{align}
\psi_3&=\pi^\frac{D}{2} \brk*{ c_1 \,
		G_{000} ^{\alpha_1 \alpha_2 \alpha_3\gamma_1\gamma_2}
		+ c_2  \,
		G_{001-\gamma_2} ^{\alpha_1 \alpha_2 \alpha_3\gamma_1\gamma_2} }, 	
\end{align}
maps to the solution given in \cite{Boos:1990rg}, if the constants 
$c_1$ and $c_2$ are chosen as
$
c_1=\Gamma_{\alpha_1}\Gamma_{1-\gamma_2}/\Gamma_{\alpha_1-\gamma_2+1}\Gamma_{\gamma_1}
$
and
\begin{equation*}
c_2=\frac{\Gamma_{\alpha_1}\Gamma_{\gamma_1-\alpha_2}\Gamma_{\gamma_1-\alpha_3}\Gamma_{\gamma_2-1}\Gamma_{2-\gamma_2}}
{\Gamma_{\gamma_1}\Gamma_{\gamma_2}\Gamma_{1+\alpha_1-\gamma_2}\Gamma_{1+\alpha_2-\gamma_2}\Gamma_{1+\alpha_3-\gamma_2}}.
\end{equation*}
In this last example we have thus successfully employed the 
level-one  symmetry to constrain the functional form of an integral without conformal level-zero symmetry. This suggests an alternative interpretation of this non-local symmetry.
\section{Massive Momentum Space Symmetry}
\label{sec:MassMomSpace}

In the massless case, the Yangian symmetry of amplitudes in planar $\mathcal{N}=4$ SYM theory \cite{Drummond:2009fd} as well as in the fishnet theory \cite{Chicherin:2017cns} can be understood as the closure of a dual (region momentum space) and an ordinary (momentum space) conformal symmetry.
It is thus a natural question whether the discovered massive Yangian can be understood as the closure of the massive dual conformal symmetry generated by \eqref{eqn:massdualconfrep} and some massive generalization of momentum space conformal symmetry. To address this question we translate the level-one generator $\levo{P}^\mu$ of \eqref{eq:DefPhat} given in terms of region momenta $x_j$ into momenta $p_j$ defined via
$p^{\mu}_j=x^{\mu}_j-x^{\mu}_{j+1}$. The masses, on the contrary, stay untouched because the mass dependence of the integrals cannot be phrased in terms of dual masses $m_j-m_{j+1}$ only. By making the respective substitutions and using the chain rule, one finds 
\begin{equation}
\label{eq:PhatToK}
\levo{P}^\mu=\ihalf \gen{\bar K}^\mu \, ,
\end{equation} 
which holds up to terms that vanish when the integral is expressed in terms of the first $n-1$ momenta by using momentum conservation. Here, the special conformal operator $\gen{\bar K}^\mu$ forms part of the following massive generalization of the momentum space conformal generators~\footnote{Here we could also use a local version of the dilatation generator density acting only on the mass $m_j$. In the tensor product this makes no difference.}:
\begin{align}
\gen{\bar P}_{j}^\mu &= p_j^\mu \, , 
\hspace{1cm} \gen{\bar L}_{j}^{\mu\nu} =  p_j^\mu \partial_{p_j}^{\nu} -  p_j^\nu \partial_{p_j}^{\mu},
\nonumber\\
\gen{\bar D}_{j} &= p_{j\nu} \partial^{\nu}_{p_j}+ \sfrac{m_j \partial_{m_j}+m_{j+1} \partial_{m_{j+1}}}{2} + \bar{\Delta}_j,
\label{eqn:MomConfMassRep}
\\
\gen{\bar K}_{j}^\mu 
&= p_j^\mu \partial_{p_j}^2
-2\brk[s]2{p_{j\nu} \partial_{p_j}^\nu+\sfrac{m_j \partial_{m_j}+m_{j+1} \partial_{m_{j+1}}}{2} + \bar{\Delta}_j }\partial_{p_j}^{\mu}.
\nonumber 
\end{align}
These generators thus furnish symmetries of the above integrals expressed in momentum space and they obey the ordinary conformal algebra. We have $\bar \Delta_j=
\half\Delta_{j}+ \half\Delta_{j+1} +  \Eval_j - \Eval_{j+1}$ with the evaluation parameters $s_j$ determined by the rules \eqref{eq:EvalRules}
\footnote{For the one-loop $n$-gons for instance, we thus find $\bar \Delta_j = a_j+a_{j+1}$.}.
We note that the operator $\gen{\bar K}_{j}^\mu$ includes a mildly non-local contribution acting on the masses of nearest-neighboring legs $j$ and $j+1$ of the Feynman graph.
In the massless limit, the above translation \eqref{eq:PhatToK} of the Yangian level-one generator to momentum space also makes connection to the momentum space Ward identities that were recently studied for conformal correlation functions, see e.g.\ \cite{Coriano:2013jba,Bzowski:2013sza,Coriano:2020ccb}.

\section{Conclusions and Outlook}

The main results presented in this letter are summarized as follows:
An infinite number of specific massive Feynman integrals features a highly constraining Yangian symmetry --- a hallmark of integrability.
The observed Yangian algebra can be understood as the closure of a novel momentum space conformal symmetry and the known massive dual conformal symmetry; this strongly suggests an embedding of these findings into a larger picture within the AdS/CFT duality.
Our finding shows that the massive IR regulator proposed in \cite{Alday:2009zm} does not only preserve dual conformal symmetry but also the Yangian symmetry of Feynman integrals.

There are plenty of further directions worth being investigated.
Firstly, it would be important to formally prove our conjecture of higher loop Yangian and massive momentum space conformal symmetry. This may be possible along the lines of the massless case \cite{Chicherin:2017cns,Chicherin:2017frs}, but it could be tricky to find a proof that applies directly to all cases considered in this letter (e.g.\ to generic spacetime dimensions $D$). 
Secondly, the applicability of Yangian symmetry to massive Feynman integrals and generic (double) $n$-gons allows us to refine the algorithm of \cite{Loebbert:2019vcj} on many more examples whose complexity interpolates between the simplest massless cases with $2$ and $9$ variables, respectively. This may finally open the door to computing e.g.\ the massless hexagon or double box integral. 
Moreover, some of the integrals discussed above have a geometric interpretation, see e.g.\ \cite{Nandan:2013ip,Davydychev:2017bbl,Herrmann:2019upk,Bourjaily:2019exo}. It would be fascinating to understand the considered symmetries in this context. 
Another highly interesting task is to seek for a massive generalization of the Basso--Dixon formula for massless fishnet four-point integrals \cite{Basso:2017jwq}, whose finding was motivated by integrability.
Finally, in the massless situation integrability took its way from AdS/CFT via the fishnet theory to Feynman integrals \cite{Gurdogan:2015csr,Chicherin:2017cns,Chicherin:2017frs}. Here we have found similar properties for the massive building blocks of quantum field theory. Tracing back these symmetries to an origin within the AdS/CFT duality
is very suggestive and may open the door to understanding integrability in massive phases of AdS/CFT.
In this direction we note that in $D=4$ for instance, the massive square fishnet diagrams discussed above can be considered as planar off-shell amplitudes generated by the Lagrangian
\begin{align}
\mathcal{L}=
&N_\text{c} \text{tr}
\brk*{
-\partial_\mu\bar X \partial^\mu X -\partial_\mu\bar Z\partial^\mu Z+\xi^2 \bar X\bar Z X Z
}
\\
&- N_\text{c}\brk*{m_j-m_k}^2 \bar Z^j{}_k Z^k{}_j
-N_\text{c}\brk*{ m_j-m_k}^2 \bar X^j{}_kX^k{}_j.\nonumber
\end{align}
Here, $N_\text{c}-n$ of the masses $m_j$ are taken to be zero, such that graphs with massless propagators in the bulk of the diagram are dominant in the planar limit~\footnote{Note that the mass terms generated by spontaneous symmetry breaking in the fishnet theory have product form instead of the difference form \cite{Karananas:2019fox,Loebbert:2020tje}.}. 
It would be interesting to test this Lagrangian for Yangian symmetry with the methods of~\cite{Beisert:2017pnr}.

\medskip 

\acknowledgments{
FL and DM are very grateful to Till Bargheer and Jan Plefka for initial collaborations on similar ideas in the context of amplitudes in $\mathcal{N}=4$ SYM theory and for helpful discussions. 
We thank Niklas Beisert, Lance Dixon and Jan Plefka
for helpful comments on the manuscript.
 The work of FL is funded by the Deutsche Forschungsgemeinschaft (DFG, German Research Foundation)--Projektnummer 363895012. JM is  supported  by  the  International  Max  Planck  Research  School  for  Mathematical  and Physical Aspects of Gravitation, Cosmology and Quantum Field Theory. DM was supported by DFF-FNU through grant number DFF-FNU 4002-00037. The work of HM has been supported by the grant no.\ 615203 of the European Research Council under the FP7 and by the Swiss National Science Foundation by the NCCR SwissMAP. 
}


\bibliography{YangianMassiveIntegrals}

\end{document}